# CiteScore metrics: Creating journal metrics from the Scopus citation index

Authors: Chris James [0000-0003-1619-5204], Lisa Colledge [0000-0002-8921-7147], Wim Meester [0000-0001-9350-3448], Norman Azoulay [0000-0003-1453-4882], Andrew Plume [0000-0002-4942-1426]

Affiliations: Chris James, Lisa Colledge and Wim Meester: Elsevier, Radarweg 29, 1043 NX Amsterdam, Netherlands
Norman Azoulay: Elsevier, Suite 800, 230 Park Ave, New York, NY 10169, United States
Andrew Plume: Elsevier, The Boulevard, Langford Lane, Kidlington OX5 1GB, UK

Corresponding author: Andrew Plume (a.plume@elsevier.com)

Highlights:

- Journal citation metrics remain important in the age of responsible metrics

- CiteScore metrics were designed on the principles of transparency, comprehensiveness and currency

- CiteScore metrics are calculated from snapshots of a dynamic database

- CiteScore Tracker is updated monthly to show progress towards the static CiteScore value as it builds

- Feedback after the initial release of CiteScore is acknowledged and resulting developments are discussed

## Abstract

In December 2016, after several years of development, Elsevier launched a set of transparent, comprehensive, current, and freely-available journal citation metrics called CiteScore metrics. Most of the CiteScore metrics are static, annual indicators calculated from the dynamic Scopus citation index. In the spirit of recent public statements on the responsible use of metrics, we outline the desirable characteristics of journal citation metrics, discuss how we decided on the cited and citing publications years and document types to be used for CiteScore metrics, and detail the precise method of calculation of each metric. We further discuss CiteScore metrics eligibility criteria and online display choices, as well as our approach to calculating static indicators from the dynamic citation index. Finally, we look at the feedback the metrics have so far received and how CiteScore is already developing in response.

## 1. Journal citation metrics in the age of responsible metrics

Journal[1] citation metrics remain relevant in the 21st century, having emerged from technological advances in the mid-20th century that permitted comprehensive citation indexing across large volumes of the published literature (Henk F. Moed, 2017). Researchers may refer to journal citation metrics when considering where to submit manuscripts, when compiling cited reference lists, and when considering which new publications are worthy of their attention and reading time (Callaway, 2016).

There have been several recent public statements around the responsible use of metrics in general, and journal citation metrics in particular. The San Francisco Declaration on Research Assessment (DORA) (Raff, 2013), the HEFCE Metric Tide report (Wilsdon et al., 2015) and the Leiden Manifesto (Hicks, Wouters, Waltman, De Rijcke, & Rafols, 2015) have laid ground rules for the appropriate use of metrics. The "2 Golden Rules" (L. Colledge, James, Azoulay, Meester, & Plume, 2017) are intended to distil the essence of these statements for non-specialists and provide a memorable guide on the responsible use of research metrics when assessing and benchmarking performance. They are the outcome of extensive engagements with a wide variety of stakeholder groups all over the world, including but not limited to researchers (as authors, peer reviewers and journal editors), bibliometricians, research office managers, research funders, university librarians and journal publishers. The first Golden Rule is to always use both qualitative and quantitative input for decisions. The second Golden Rule is to always use more than one research metric as the quantitative input, and this guidance is reflected in all three of the aforementioned public statements.

The second Golden Rule acknowledges that performance cannot be expressed by any single metric, as well as the fact that all metrics have specific strengths and weaknesses. Therefore, using multiple complementary metrics can help to provide a more complete picture and reflect different aspects of research productivity and impact in the final assessment. For example, field-weighted citation impact (FWCI), a Snowball Metric (Snowball Metrics Recipe Book, 2017) which is the most often-selected metric in SciVal (see scival.com), is a size-independent indicator applicable to entities at all levels of aggregation from author, institution and country to the topic of research and the journal. FWCI for a given entity is the average of the ratio of citations that each publication associated with that entity has received relative to the expected number of citations for publications of the same subject, document type and year of publication (Snowball Metrics Recipe Book, 2017). This normalisation allows direct comparison of different entities with each other and intrinsically benchmarks against the world average. As with all metrics, it has its drawbacks: the metric gives no indication as to the number of citations that an entity has received and is computationally (if not conceptually) complex and not amenable to checking by hand.

As such, pairing FWCI with a simpler size-independent metric – such as average citations per publication, which is also a Snowball Metric – gives a complementary view on performance. Average citations per publication is simple to calculate from appropriate base data and amenable to hand-checking, but unlike FWCI, is not directly comparable across subject areas, time and entity sizes. Adding complementary metrics – including size-dependent ones such as total citations, or non-citation-based ones such as online usage or attention – can offset their respective shortcomings and provide a richer, multi-dimensional understanding of the entity of interest.

---

[1] Throughout this article, we use the term 'journal' in a broad sense. Owing to the wide coverage of the Scopus citation index across serial (journals, book series, conference series and trade journals) and non-serial (e.g. stand-alone conference proceedings and books) peer-reviewed literature, we use the term 'journal' here to include all serial peer-reviewed publication venues for which citation metrics can be calculated. See section 2.4 for more details.

## 1.1. What types of journal citation metrics are currently available?

Currently the most well-known journal citation metric is the Journal Impact Factor (JIF), calculated by Clarivate Analytics (formerly the IP & Science division of Thomson Reuters). The JIF has been used for over 40 years by researchers, librarians and publishers as an indicator of a journal's citation impact (Larivière & Sugimoto, 2018). Simply put, the JIF is an indicator of average citations per publication in a journal over a 2-year time window (Archambault & Larivière, 2009). The June 2018 release of the Journal Citation Reports (JCR) included 11,655 journals and was based on an extract of the Web of Science database (Clarivate Analytics, 2018).

The JIF has received criticism within the research community, for the lack of transparency of the metric (Archambault & Larivière, 2009; PLoS Medicine Editors, 2006) and for the potential for manipulation (Martin, 2016; Matthews, 2015; Vanclay, 2012). Recently Clarivate Analytics have tried to address these issues by highlighting several of the supporting metrics in the JCR, including the Average Journal Impact Factor Percentile, Normalized Eigenfactor and Percent Articles in Citable Items. In the 2018 release of JCR, further improvements were made with the inclusion of the citation distribution across publications included in the JIF calculation (Clarivate Analytics, 2018). In response to the JIF criticism, the Nature Publishing Group journals now includes the two-year median of citations calculated by the publisher (and taken not from JCR data but from the article-level Web of Science dataset), to give an additional view of their journals' performance (Nature Research Journal Metrics webpage, https://www.nature.com/npg_/company_info/journal_metrics.html).

In 2010, two journal citation metrics were launched by independent bibliometric groups. Source-Normalised Impact per Paper (SNIP) and Scimago Journal Rank (SJR) are calculated each year using Scopus data by the Centre for Science and Technology Studies (CWTS) at Leiden University and Scimago Lab respectively and cover more than 22,000 serial titles (Lisa Colledge et al., 2010). Both metrics employ sophisticated approaches to field-normalisation, addressing one of the most common criticisms of the JIF. However, the drawback of both metrics is that they are algorithmically complex and so are not transparent or trivial to replicate. After the launch of SNIP and SJR the research community had access to sophisticated metrics to assess journal citation impact in Scopus, but no simple and easily-replicable measure of citation impact.

## 1.2. What are the desirable characteristics of journal citation metrics?

To complement the sophisticated SNIP and SJR metrics, new citation metrics included in the Scopus database needed to possess several desirable characteristics. They needed to:

- Be **transparent** in terms of the method of calculation and the source of the underlying data

- Be **comprehensive** in terms of being available across the broadest possible coverage of serial scholarly publications including journals, book series and conference proceedings

- Be **current** in terms of being calculated as soon as possible after the underlying data are deemed sufficiently complete for reliable measurement, and within a calculation framework of defined citing and cited time periods (publication years) that provides a degree of stability from one calculation period to the next

These considerations were accounted for when creating the new set of journal metrics that comprise CiteScore metrics, which are made freely available alongside SNIP and SJR in acknowledgement of the second Golden Rule. Taken together, the characteristics noted above make CiteScore metrics a fitting complement to more sophisticated metrics such as SNIP and SJR within the Scopus citation index.

## 2. Designing CiteScore, the journal citation metric at the core of CiteScore metrics

To build on the broad familiarity that researchers have with other size-independent journal citation metrics, CiteScore was designed from the beginning as the central element of CiteScore metrics, from which the other CiteScore metrics would be derived. In this section we describe the decision-making process behind the design of CiteScore metrics, including CiteScore around the citing and cited publication years, citing and cited document types, the method of calculation of CiteScore metrics, eligibility criteria for Scopus-covered journals to have CiteScore metrics calculated, and design choices in how to display CiteScore metrics. Table 1 summarises the parameters used by existing journal metrics.

| Journal citation metric | Source database | Citing publication year | Cited publication year | Citing document types included (for citation counts) | Cited document types included (for size normalisation) |
|---|---|---|---|---|---|
| SNIP | Scopus | 1 year (Y) | 3 years (Y-1, Y-2, Y-3) | Citations from articles, reviews, and conference papers | Article, review, conference paper |
| SJR | Scopus | 1 year (Y) | 3 years (Y-1, Y-2, Y-3) | Citations from articles, reviews, and conference papers | Article, review, conference paper |
| Journal Impact Factor (JIF) | JCR* | 1 year (Y) | 2 years (Y-1 and Y-2) | Citations from all indexed publications to indexed journals (not necessarily linked to specific publications) | "Citable items" including article, review, conference paper (McVeigh & Mann, 2009) |
| 5-Year Journal Impact Factor | JCR* | 1 year (Y) | 5 years (Y-1, Y-2, Y-3, Y-4, Y-5) | Citations from all indexed publications to indexed journals (not necessarily linked to specific publications) | "Citable items" including article, review, conference paper (McVeigh & Mann, 2009) |

*Table 1: Summary of some existing journal citation metrics and some of their parameters. * These metrics are not calculated from Web of Science, but from the JCR analytical database extracted annually from Web of Science and which relies not on article-based but journal-based citation matching (Hubbard and McVeigh, 2011). This underlying database is not available to users.*

### 2.1. Citing and cited publication years

Each of the metrics shown in Table 1 uses a single citing publication year and a multi-year cited publication period. For CiteScore, the same single citing publication year was chosen because it maintains currency and is most responsive to changes by focussing on recent citation activity and is also familiar and consistent with these existing journal metrics.

The cited publication years for SNIP and SJR, which are both based on Scopus data, are the 3 years prior to the citing publication year. This is seen as the best compromise for a broad-scope database, incorporating a reasonable proportion of lifetime citations in a majority of disciplines, while still reflecting relatively recent activity (Lisa Colledge et al., 2010; Lancho-Barrantes, Guerrero-Bote, & Moya-Anegón, 2010). The evidence in favour of using a 3-year cited publication period that led to the selection of a 3-year window for SNIP and SJR was also used as the basis for selecting this option for CiteScore.

## 2.2. Citing and cited document types

Journals exist as wrappers for a diversity of different material that is published under a variable and discipline-specific set of rubrics, ranging from original research contributions and literature reviews through to short surveys, letters to the editor, and technical readouts. All such publications could be cited and may themselves contain cited references. Any arbitrary reduction of such a rich literature into a binary 'citable' or 'non-citable' designation is fraught with difficulties, and maty have unintended consequences. For example, the oft-criticised use of a 'citable item' count for size-normalisation in the JIF calculation (whilst the numerator counts all citations to the *journal*, not to specific *publications*, and places no document type restriction on the citing publications) can be avoided by including all indexed publications in the denominator. The situation in which editors and publishers may lobby to exclude certain types of publications from the denominator to help increase the JIF (Martin, 2016) would no longer exist, as would the loophole whereby journals publishing "non-citable items" which are nonetheless cited are at an advantage, *ceteris paribus*, over journals that do not publish such material. While SNIP and SJR limit the document types for both the cited and citing publications to articles, reviews and conference papers, the classification of documents into these document types is also somewhat subjectively applied and can lead to a lack of transparency and trust in the metric (Davis, 2016).

For the greatest possible transparency, CiteScore includes all document types indexed in Scopus for both the citing and cited publications, except for articles-in-press because their publication year is not yet finalised, and they do not yet include references and therefore can be cited but cannot give citations. This decision recognises the overall influence of a journal as a citing and cited entity. Since any publication can be cited (and any publication containing cited references can be a source of citations to other publications), it follows that the overall influence of a journal as a citing and cited entity is most fully reflected by including all its publications in the citing and cited dimensions.

## 2.3. CiteScore and CiteScore metrics calculation methods

Based on the above decisions, the CiteScore calculation method is as follows for the example of the 2017 calculation (see Figure 1):

Citations (A) is the sum of citations in 2017 (i.e. where the citing publication year is 2017) that are linked to publications in a journal with a (cited) publication year of 2014, 2015 or 2016.

Documents (B) is the count of all publications in a journal with a publication year of 2014, 2015 or 2016, excluding articles-in-press.

CiteScore is calculated by dividing the Citations (A) by Documents (B). CiteScore is calculated to two decimal places as a trade-off between minimising tied rankings and the statistical precision arguments of the Leiden Manifesto (Hicks, D., Wouters, P., Waltman, L., De Rijcke, S., & Rafols, I. (2015)).

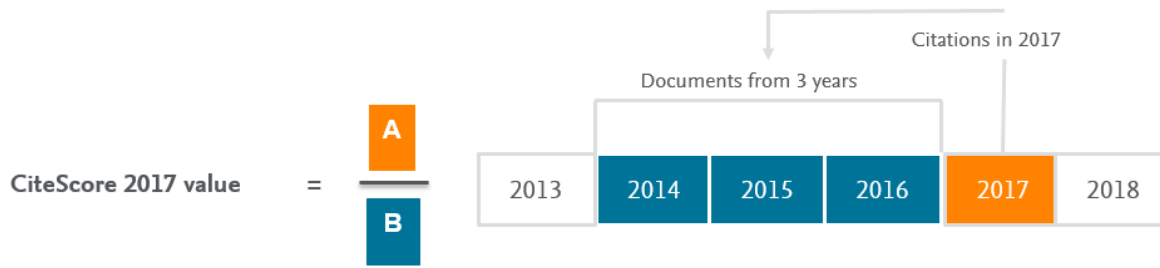

*Figure 1: The method used to calculate 2017 CiteScore.*

As noted above, a set of complementary CiteScore metrics can be derived from CiteScore itself in recognition of the second Golden Rule. These additional metrics help contextualise CiteScore and permit a degree of cross-subject comparison if required. Table 2 details the "basket" of additional metrics calculated using the CiteScore calculation framework.

| Metric | Contribution to the basket of metrics |
| --- | --- |
| **CiteScore** | A size-independent journal citation metric indicating mean citations per publication. |
| **CiteScore Percentile** | Percentile rank by descending CiteScore within the journal's assigned ASJC subject category (or categories)[2], allowing comparison of relative CiteScore performance between journals with different subject classifications. |
| **Documents ("B" in Figure 1)** | An indication of the size of the journal, and equal to the CiteScore denominator. |
| **Citations ("A" in Figure 1)** | An indication of total or size-dependent citation impact of the journal, and equal to the CiteScore numerator. |
| **CiteScore Rank and Rank out of (n in category)** | Rank by descending CiteScore within the journal's assigned ASJC subject category (or categories)[2]. The Rank out of (n in category) gives an indication of CiteScore rank relative to category size. |
| **CiteScore Quartile** | Quartile rank by descending CiteScore within the journal's assigned ASJC subject category (or categories)[2], offering a broad indication of relative CiteScore performance. |
| **Percent (%) Cited** | Proportion of documents (i.e. the CiteScore denominator) that have received at least one citation contributing to the sum of citations (i.e. the CiteScore numerator). This gives an indication of the reliance of CiteScore on a greater or lesser proportion of the publications appearing in the journal. |

*Table 2: CiteScore and derived CiteScore metrics.*

---

[2] In Scopus, journals are classified in one or more subject categories using the All Science Journal Classification (ASJC).

CiteScore Percentile for a serial title with CiteScore value of X, is calculated by applying the equation:

CiteScore Percentile for a serial title with CiteScore value of X = [ (L + (0.5 x S) ) / N ] x 100

where:

L = number of journals in the ASJC subject field with a CiteScore lower than X

S = number of journals in the ASJC subject field with CiteScore X

N = total number of journals in the ASJC subject field with a CiteScore

As such, a journal with a CiteScore Percentile of 96% is ranked by CiteScore as high or higher than 96% of journals in that category. This method does not give a 100th percentile rank and the percentile rank is rounded down to the nearest whole percentage, in line with the spirit of this calculation method. Note that a journal has one CiteScore, but a different CiteScore Percentile for each ASJC subject field it is indexed in.

CiteScore Quartiles are derived from CiteScore Percentiles and are defined as Quartile 1 (99-75th percentiles), Quartile 2 (74-50th percentiles), Quartile 3 (49-25th percentiles) and Quartile 4 (24-0th percentiles). As per CiteScore Percentile, a journal may have a different CiteScore Quartile for each ASJC subject field it is indexed in.

## 2.4. CiteScore metric eligibility criteria

There are several criteria for determining the eligibility of a journal to have CiteScore metrics:

- The journal must be selected for indexing in Scopus. Scopus has an active Content Selection and Advisory Board (CSAB) made up of independent subject experts who evaluate submitted titles throughout the year ("Content – Scopus – Solutions | Elsevier,").

- The journal must be actively indexed in Scopus and classed as a journal, book series, trade journal, or serial conference proceedings. Stand-alone books or conference proceedings are not eligible.

- The journal must have at least one publication in the 3-year cited publication period (i.e. the Documents count that forms the denominator of CiteScore).

## 2.5. Displaying CiteScore metrics

To make CiteScore metrics not only transparent, comprehensive and current but also *accessible*, considerable attention was given to how to display the metrics in a free-to-access online environment. It was recognised from looking at rankings such as the Scimago Journal Rank (http://www.Scimagojr.com/journalrank.php), CWTS Journal Indicators (http://www.journalindicators.com) and the Times Higher Education World University Rankings (https://www.timeshighereducation.com/world-university-rankings/2018/world-ranking), that users of such rankings want to search, sort and export information for further sharing and reuse. The logical location for such functionality for CiteScore was Scopus, which is a subscription-based product. The ideal solution would be to offer a "free layer" of Scopus that would allow easy access to those pages that would house CiteScore metrics, but this required significant development and would have delayed the initial release of CiteScore.

To offer the functionality in a timely manner, a separate website was developed that was "powered by Scopus", linking seamlessly into Scopus.com and providing all the functionality users would expect. The resulting site, https://journalmetrics.scopus.com, was developed by a third-party supplier. This was subsequently migrated to the newly-developed "free layer" of Scopus in late May 2018.

However, CiteScore metrics were never designed to reside in only one place, so since the launch in December 2016 CiteScore metrics have also been available in SciVal (also powered by Scopus), Elsevier journal homepages (e.g. https://www.journals.elsevier.com/cell-stem-cell) and journal insights pages (e.g. https://journalinsights.elsevier.com/journals/0142-9612). Shortly after launch CiteScore metrics were also available in Pure (https://www.elsevier.com/solutions/pure) and other publisher websites (e.g. http://www.emeraldgrouppublishing.com/aaaj.htm and https://www.tandfonline.com/toc/cpro20/current). They can be incorporated into other websites easily, for example for researchers, institutions and publishers to showcase the quality of the titles with which they work, according to these guidelines (https://www.elsevier.com/__data/assets/pdf_file/0011/239078/CiteScore_TermsAndConditions.pdf).

## 3. Hitting a moving target: Calculating static metrics with a dynamic citation index

Metric values for a given journal are often compared over time to give a sense of performance development, and the case of journal citation metrics is no different. Each year, many journal editors and publishers rush to promote their new JIF values and compare them to the previous year's results. The desire for comprehensiveness and currency must be traded off against the realities of a dynamic citation index, which is subject to ongoing addition of newly-covered journals and the backfilling of previously-published content, as well as changes resulting from error corrections or incorporation of missing articles or issues. If the metrics designer wants the data underlying these metrics to be transparent, it presents a simple choice: to create static metrics calculated on a snapshot of the source database at a specific point in time each year, or to recalculate all historical metrics afresh. For example, a static snapshot of the Scopus database taken in May 2017 might be used to calculate annual 2016 CiteScore metrics, and another snapshot taken in May 2018 might be used to calculate annual 2017 values. The advantage of this approach would be that the 2016 CiteScore metrics are immutable and permanent and would also be directly comparable with the 2017 CiteScore metrics calculated 12 months later.

This approach neglects to account for fresh content that may have been added and errors corrected in the Scopus data CiteScore metrics that could now be reflected in earlier years of CiteScore metrics if they were to be recalculated using the latest snapshot. However, indexing lags (i.e. time elapsed from publication to indexation, which may vary across journals, publishers, subjects and countries) might lead to systematic bias against the most recent publication year when recalculating historical metrics, making trend comparisons unreliable.

While the sophisticated SNIP and SJR metrics use the latter approach of historical recalculation of earlier metrics from a fresh data snapshot each year, the effect is far less visible to the end-user than with much simpler metrics as JIF, which employs the static snapshot approach and produces immutable and permanent metrics each June (except for a restatement in September for a small number of journals omitted or with obvious errors). Keeping in mind the principles of transparency, comprehensiveness and currency discussed above, the static snapshot approach was deemed the best option for CiteScore metrics.

## 3.1. Indexing lags at Scopus: when to snapshot?

In large citation indexes such as Scopus, there is always a period that elapses between an article being published (generally regarded as its first appearance on the journal website) to it being visible in Scopus. This period can be as little as a few days, or as great as many months. On average, content passed electronically by the publisher to Scopus is processed and available at Scopus.com within 5 days. Additional elapsed time from publication to indexing may be due to intermittent data provision, missing or undeliverable content, and so on. For example, many of the larger publishers have sophisticated electronic delivery mechanisms which feed content to Scopus continuously, whereas for some publishers' electronic content is downloaded in batches. Moreover, about 5% of the content submitted to Scopus is in the form of physical printed issues, which must be digitised and enriched prior to the regular indexing process. These workflow considerations are in addition to the fact that some journals may fall behind their publication schedule and Scopus must wait for issues beyond the announced publication date.

Taking these issues into account, indexed content for the most recent publication year continues to be added to Scopus beyond the end of the calendar year. The desire for comprehensiveness must be balanced against the need for currency, and so we conducted an analysis to look at what point in the year following the end of a publication year that CiteScore (and hence the other CiteScore metrics) had reached a stable value for most journals. Our analysis of CiteScore for any given recent year showed that after steep growth as the numerator (citation count) builds month-on-month throughout the citing publication year, it reaches a high and stable level as early as January and not later than April of the following year; consequently, CiteScore metrics are calculated on a static snapshot of the citation index taken in May. We observed however that the incomplete CiteScores calculated in the early months of the year (i.e. prior to April) nonetheless offered a stable view of the citation impact of journals relative to their peers.

## 3.2. CiteScore Tracker: a dynamic metric to complement a static one

Because the numerator (citation count) builds month-on-month throughout the citing publication year, we decided to create one further CiteScore metric, called CiteScore Tracker. CiteScore Tracker is calculated in the same way as CiteScore, but for the current (citation) year rather than previous, complete years; it is not based on a rolling 12-month window of citing publications but builds a partial view over time as indexation continues. The CiteScore Tracker is updated every month to reflect citations from newly-indexed publications and is offered as a current indication of a title's performance, especially in relative terms.

For example, in May 2018 the CiteScore metrics for 2017 were released as static values. From June 2018, CiteScore Tracker values were made available to end-users based on yet-incomplete 2018 citing publication data and (almost complete) 2015-2017 cited publication data. Of course, as a metric with a partially-complete numerator these CiteScore Tracker values typically start lower than the previous year's static CiteScore values. Nonetheless, users can return throughout the year as the citation index moves towards completion early the following year, right up to the point where a fixed snapshot of the database is taken for the calculation of the fixed, annual CiteScore metrics.

The benefits of CiteScore Tracker to users were explored through two series of one-to-one user tests in April and October 2016 with 6 participants each time, consisting of researchers, librarians and publishers. Interviews were held by Scopus' User Experience team via telephone and screen sharing. Each interview took around an hour and the results were reported anonymously, so that each participant could speak freely and honestly.

Participants indicated that for established journals, CiteScore Tracker is useful for monitoring the development of CiteScore (c.f. the previous static annual value) throughout the year, and for benchmarking CiteScore Tracker for a given journal against other journals in the field. These use cases are not just of interest to editors and publishers, but also to readers and (potential) authors when evaluating their reading and manuscript submission priorities. Participants also liked CiteScore Tracker because of its ability to provide a newly-launched or newly-indexed journal with metrics in a relatively short period.

According to Clarivate, a new (or newly-indexed) journal can only receive a JIF after 3 complete indexed years of source data (Clarivate Analytics, 2017b). That means a title which started publishing in 2017 will receive a JIF around June 2020, up to 3.5 years after it started publishing. For CiteScore, Scopus begins to calculate a metric for the title as soon as it has one year of publications to contribute towards the denominator. This means for a journal that started publishing in 2017 and was promptly accepted for Scopus indexing, it could receive its first annual CiteScore (CiteScore 2018) not later than June 2019. With CiteScore Tracker however, an indication of the average citation impact of publications in this journal would be available by May 2018, not more than 1.5 years after it started publishing (see Table 3).

|  | CiteScore Tracker (calculated monthly) | Annual (static) CiteScore | 2-year JIF |
|---|---|---|---|
| Time required | Between 0.5 and 1.5 years | Between 1.5 and 2.5 years | 3.5 years |

*Table 3: Time required for a new (or newly-indexed) journal to receive journal citation metrics.*

### 3.3. Writing history in the present day: creating static snapshots of Scopus to calculate historical CiteScore metrics

To provide a suitable view of historical trends for CiteScore metrics at the time of the initial launch, an approach to retrospectively calculating the metrics was devised that simulated the snapshots required for the calculation of the static annual values. Put another way, we had to recreate a snapshot of Scopus content at a given date in history from our reference point in around mid-2016: the following elucidates how this was achieved.

As a dynamic citation index, Scopus processes thousands of publications per day and – along with hundreds of other metadata tags - each new publication is stamped with a "load date" (i.e. the date of entry of the publication record into the index) and a "sort year" (i.e. the year of publication). The load date remains in the metadata of the record in perpetuity, even if adjustments are made to the metadata of the article later. Using the load date, we could identify when the records were loaded in Scopus for each year and recreate Scopus as it existed on certain dates in previous years (see Table 5). In practice this meant for CiteScore metrics 2011 (for example), the publications used in the calculations were loaded into Scopus before 31st May 2012.

| CiteScore year | Cited publication years | Citing publication year | Load date up to |
|---|---|---|---|
| 2017 | 2014, 2015, 2016 | 2017 | 30 April 2018 |
| 2016 | 2013, 2014, 2015 | 2016 | 31 May 2017 |

| 2015 | 2012, 2013, 2014 | 2015 | 31 May 2016 |
| 2014 | 2011, 2012, 2013 | 2014 | 31 May 2015 |
| 2013 | 2010, 2011, 2012 | 2013 | 31 May 2014 |
| 2012 | 2009, 2010, 2011 | 2012 | 31 May 2013 |
| 2011 | 2008, 2009, 2010 | 2011 | 31 May 2012 |

*Table 4: The publication year ("sort year") and load date used for each CiteScore metrics calculation between 2011 – 2017. Owing to improvements in processing speeds, the snapshot for the 2017 calculation was taken a month earlier than in previous years.*

## 4. Initial feedback after the CiteScore metrics release

The initial launch of CiteScore metrics on 8th December 2016 was accompanied by significant promotional activity and attracted a great deal of attention in blogs and social media channels such as Twitter. Within 2 weeks there had been some 3,000 tweets about CiteScore metrics, around 1,300 of which appeared within the first 2 days after launch. Within the first year since launch we noted at least 17 journal publishers that display CiteScore metrics on their websites, showing a clear appetite for the new set of journal metrics. The initial reactions were a mixture of positive and negative, with several themes surfacing about document type classification, potential Elsevier bias and conflict of interest, and whether there was a need for another journal citation metric.

### 4.1. Document type classifications and potential Elsevier conflict of interest

The fact that the CiteScore calculation includes all document types in both the cited and citing publications received both praise and criticism (L.Waltman, 2017). This inclusiveness was welcomed for the fact that it gives a complete and consistent view with respect to both the numerator and denominator of CiteScore, which is not the case for the JIF (Martin, 2016). This in turn makes the metric more transparent, easier for end-users to check for themselves, and removes one of the more common loopholes by which journal citation metrics can be gamed.

On the other hand, including publications of all document types in the calculation of CiteScore was also criticised as problematic, because it is seen to "penalise" titles like Nature and The Lancet. Such journals historically publish a significant volume of front matter content (editorials, notes, letters to the editor, etc.) which typically are not well-cited (if at all). Conversely, this front matter content is excluded from the JIF denominator, whilst any citations this content may receive is included in the numerator. For this reason, the JIF of such journals will often be greater than the CiteScore. We argue that this is not a shortcoming of either metric, but simply an emergent property stemming from the metric design choices. Indeed, in the period when the JIF first rose to popular prominence in the 1990s, the method of excluding certain document types from the denominator was criticised as being misleading (H. F. Moed & van Leeuwen, 1996). In future, we remain open to supplementing CiteScore metrics with variants which do apply document-type restrictions.

Starting on the day of the initial CiteScore metrics release, Bergstrom and West from the Eigenfactor Project conducted an analysis, which spread over 2 weeks and 7 individual blog posts on their website (Bergstrom & West, 2016). One of the findings of their analysis was that Nature-branded journals had 2015 CiteScore values some 40% lower than the 2015 JIF; the authors might equally have claimed based on the same analysis that Nature-branded journals have 2015 JIFs that overstate their 2015 CiteScores by 29% (the inverse calculation). While this may be true (and partly influenced by different cited publication periods, coverage of citing journals and citation matching precision and recall between the two underlying citation indexes), it is also true of any journal publishing significant volumes of material not included in the JIF denominator. This includes (as the authors demonstrated) the Lancet journal family, an imprint of Elsevier. Other analysis presented in the blog series relates to changes in journal rank under JIF and CiteScore, but which does so by rank across the entire set of matched journals (rather than within discrete subject fields); where publishers have very different subject-based portfolios the rank comparison between JIF and CiteScore may be only slightly different, but still appear as massive shifts when analysed in the way presented in the blog post.

Throughout these posts was the suggestion of a potential conflict of interest represented by having a journal publisher producing a journal citation metric, and whether CiteScore metrics benefit Elsevier's publishing interests. Other blog posts and tweets raised similar concerns (Enago Academy, 2017; Philip M. Davis, 2016; Straumsheim, 2016). Initially, Bergstrom and West stated that Elsevier journals received a "boost" of 25% under CiteScore relative to what they would expect given their JIF scores. That figure was subsequently lowered to 12% once the Lancet journals were correctly included in the analysis and downgraded again to 10% by the end of the blog series. It is important to note that the authors argued from the position of an Elsevier advantage despite their statement that they did not believe Elsevier was manipulating the metric values or acting dishonourably, especially after the authors themselves pointed out that some other publishers, including Annual Reviews, IEEE, and Emerald also fared comparatively well under CiteScore versus JIF.

Finally, in a 2017 webinar organised by the Association of Learned and Professional Society Publishers, Dr Ludo Waltman said that ideally the world's largest scientific publisher should not also be the producer of one of the world's most visible journal metrics (L. Waltman & James, 2017). However, Waltman went on to say that CiteScore (and the complete set of CiteScore metrics) are transparent and freely available, so end-users can understand how the metrics are calculated and how their definition may affect the ranking of certain journals and publishers. He concluded by saying that in practice, the question of Elsevier's potential conflict of interest is not a major concern.

## 4.2. The need for another journal metric and comparison to JIF

Upon initial launch of CiteScore metrics, there were also several comments on Twitter and elsewhere regarding the need for "another journal metric". We answer this by noting that CiteScore is the world's first truly transparent, comprehensive and current journal citation metric that is easy to understand and freely available to end-users. It is based on the same underlying citation index from which more sophisticated journal citation metrics such as SNIP and SJR metrics are derived, but it uses a familiar calculation framework. These features directly answer the challenge laid out by second Golden Rule to always use multiple, complementary metrics as quantitative input in research evaluations. Moreover, the comprehensiveness of coverage of the Scopus database means that CiteScore metrics are available for journals not previously well-served with simple journal metrics. For example, analysis of the 2016 CiteScore metrics release found that 216 journals from 70 different publishers in the top 10% in their subject category by CiteScore lacked a JIF (https://www.elsevier.com/about/press-releases/science-and-technology/citescore-shines-a-spotlight-on-high-quality-journals-neglected-by-other-metrics). This number included 49 journals from Taylor & Francis, 28 from Elsevier, 20 from Springer Nature and 13 from Wiley; open access publishers, university presses and society publishers were also represented in the list.

At launch, CiteScore metrics were available for 22,256 journals, while 11,365 journals were listed with a 2015 JIF. Figure 2 shows a scatter plot of the subset of 11,079 journals we were able to match with both 2015 CiteScore and 2015 JIF values showing a coefficient of determination (r-squared) for the linear regression of around 0.75. This suggests that the two journal citation indicators are well-correlated for journals where both are available, with just 25% of the variability not explained in terms of the value of the other.

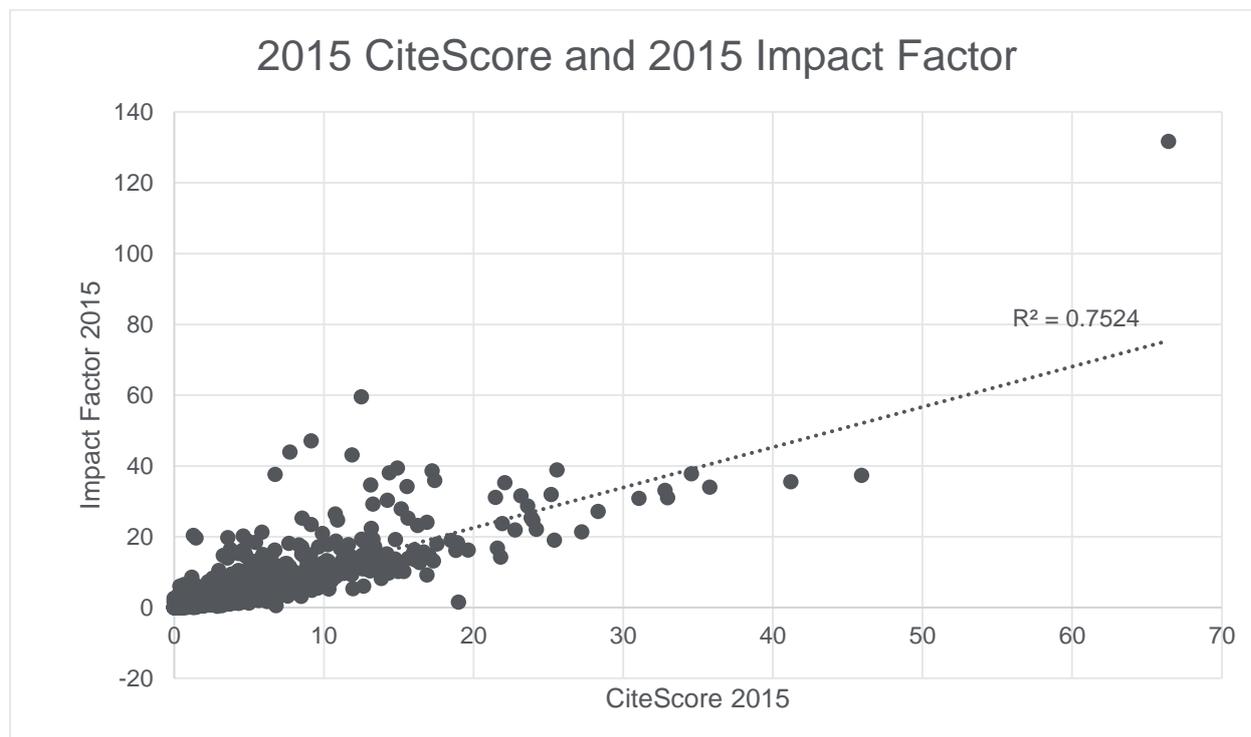

*Figure 2: CiteScore 2015 values and JIF 2015 values compared*

## 5. Recent and forthcoming developments for CiteScore metrics

Recognising that the demand for and use of metrics in research reflect the underlying mechanisms of research evaluation, which in turn reflect the management and culture of the research enterprise. As in wider society, 'culture' in research is diverse and varies across geographical and disciplinary boundaries and exists in perpetual motion. As such, CiteScore metrics have been purposefully designed to be adaptable to future requirements for simple citation metrics. For example, the calculation framework employed can be easily applied to other units of aggregation other than the journal: CiteScore metrics can be calculated for any grouping of publications connected with entities such as topics, research funders and institutions or the output of entire countries.

Since the initial release of CiteScore in December 2016 we have continued to expand the number of titles that receive a CiteScore, which from the 2015 to the 2017 metrics increased by over 1,100 journals. Some of those additions came from new coverage in Scopus, and some were a result of an evaluation to include more conference proceedings in the set of titles that receive CiteScore metrics. As Scopus continues to expand, so will the list of titles that receive CiteScore metrics.

Before the launch of the 2016 CiteScore metrics in June 2017, we implemented a change in our calculation approach for title name changes. When a title changes its name, it receives a new source ID (a unique numeric identifier) in Scopus, along with a relationship link between the former and current titles. Starting with the CiteScore metrics 2016 calculations we updated the calculation method to include all publications in the cited publication period linked to the current tile, whether published under the current or any previous title(s). This update ensures that journals changing name retain a complete view of their performance in CiteScore metrics. This is in direct contrast with the handling of journal title changes for the JIF, in which most title changes result in a 'split' JIF for the former and current titles which may confuse end-users and cause editors and publishers to avoid updating journal names as the research communities they serve evolve over time.

The increased degree of scrutiny of the underlying data in the Scopus citation index since the release of CiteScore metrics (especially from journal editors and publishers, but also from end-users) has also meant that we have been provided with helpful feedback regarding subject area classifications, publisher names and imprints, and publication and citation counts. All such feedback is directed via the in-product support function to a feedback form and data amendments and updates are processed as quickly as possible. In the 2 months following the release of the annual static CiteScore metrics values, it is also possible to request corrections to these via the same process; any appropriate recalculations are released in the second half of the year.

In future, and only after extensive testing and evaluation, we may create CiteScore metrics that do split out different document types, as this is an oft-discussed feature of other journal citation metrics.

## 6. Conclusions

From the time of its launch, there has been a significant interest in CiteScore metrics from the research community and other stakeholders, including other metrics providers. Even if in no other way, we would consider CiteScore metrics a success because it has refocused the conversation around journal citation metrics on the key themes of transparency, comprehensiveness and currency, and the relationship between research metrics, research evaluation and trust. From a technical standpoint, we have succeeded in showing how static annual metrics can be calculated from a dynamic citation index, supplemented with the dynamically-building CiteScore Tracker.

Funding: This research did not receive any specific grant from funding agencies in the public, commercial, or not-for-profit sectors.

Archambault, É., & Larivière, V. (2009). History of the journal impact factor: Contingencies and consequences. *Scientometrics*, *79*(3), 635–649. https://doi.org/10.1007/s11192-007-2036-x

Bergstrom, C. T., & West, J. (2016). Comparing Impact Factor and Scopus CiteScore, 1–11. Retrieved from http://eigenfactor.org/projects/posts/citescore.php

Callaway, E. (2016). Publishing elite turns against impact factor. *Nature*, *535*(14 July), 210–211. https://doi.org/10.1038/nature.2016.20224

Clarivate Analytics. (2018). *Clarivate Analytics Releases Enhanced 2018 Journal Citation Reports Highlighting the World's Most Influential Journals*. Retrieved from https://clarivate.com/blog/news/clarivate-analytics-releases-JCR-2018

Clarivate Analytics. (2017b). Journal Citation Reports : A Primer on the JCR and Journal Impact Factor, 8. Retrieved from http://stateofinnovation.com/journal-citation-reports-a-new-primer

Colledge, L., de Moya Anegón, F., Guerrero Bote, V., López Illescas, C., El Aisati, M., & Moed, H. (2010). SJR and SNIP: two new journal metrics in Elsevier's Scopus. *Serials: The Journal for the Serials Community*, *23*(3), 215–221. https://doi.org/10.1629/23215

Colledge, L., James, C., Azoulay, N., Meester, W., & Plume, A. (2017). CiteScore metrics are suitable to address different situations – A case study. *European Science Editing*, *43*(2). https://doi.org/10.20316/ESE.2017.43.003

Content – Scopus – Solutions | Elsevier. (n.d.). Retrieved November 24, 2017, from https://www.elsevier.com/solutions/scopus/content

Davis, P. (2016). Can Scopus Deliver A Better Journal Impact Metric? *The Scholarly Kitchen*. Retrieved from https://scholarlykitchen.sspnet.org/2016/03/07/can-scopus-deliver-a-better-journal-impact-metric/

Editors, Pl. M. (2006). The impact factor game: It is time to find a better way to assess the scientific literature. *PLoS Medicine*, *3*(6), 0707–0708. https://doi.org/10.1371/journal.pmed.0030291

Enago Academy. (2017). CiteScore: Accessible Analytics or Misleading Metrics? *Academic Writing*, 1–4. Retrieved from https://www.enago.com/academy/citescore-accessible-analytics-or-misleading-metrics/

Hicks, D., Wouters, P., Waltman, L., De Rijcke, S., & Rafols, I. (2015). Bibliometrics: The Leiden Manifesto for research metrics. *Nature*, *520*(7548). https://doi.org/10.1038/520429a

Hubbard, S.C. and McVeigh, M.E. (2011) Casting a wide net: the journal impact factor numerator. *Learned Publishing*, 24: 133–137.

Lancho-Barrantes, B. S., Guerrero-Bote, V. P., & Moya-Anegón, F. (2010). What lies behind the averages and significance of citation indicators in different disciplines? *Journal of Information Science*, *36*(3), 371–382. https://doi.org/10.1177/0165551510366077

Larivière, V., & Sugimoto, C. R. (2018). The Journal Impact Factor: A brief history, critique, and discussion of adverse effects. *Springer Handbook of Science and Technology Indicators*, (2018), 1–33. Retrieved from https://arxiv.org/pdf/1801.08992.pdf%0Ahttp://arxiv.org/abs/1801.08992

Martin, B. R. (2016). Editors' JIF-boosting stratagems – Which are appropriate and which not? *Research Policy*, *45*(1), 1–7. https://doi.org/10.1016/j.respol.2015.09.001

Matthews, D. (2015). Journal impact factors 'no longer credible.' Retrieved February 1, 2017, from https://www.timeshighereducation.com/news/journal-impact-factors-no-longer-credible

McVeigh, M. E., & Mann, S. J. (2009). The journal impact factor denominator: Defining citable
15 | P a g e